\documentclass[final,1p,times, amssymb]{elsarticle}

\journal{Materials Letters}
\usepackage{epsfig}
\usepackage{graphicx}%
\usepackage{epstopdf}
\usepackage{dcolumn}
\usepackage{bm}
\usepackage{ifthen}
\usepackage{booktabs}
\usepackage{SIunits}
\usepackage{ulem}
\usepackage{xcolor}
\usepackage{amsthm}

\begin{document}

\begin{frontmatter}





%
\title{Effect of annealing on the magnetic properties of zinc ferrite thin films}

\author[add1,add2]{Yogesh Kumar}
\ead{lucky1708@gmail.com}
\author[add1]{Israel Lorite}
\author[add1]{Michael Lorenz}
\author[add1]{Pablo Esquinazi}
\author[add1]{Marius Grundmann}
\address[add1]{Felix-Bloch Institute for Solid State Physics, Fakult\"{a}t f\"{u}r Physik und Geowissenschaften, Linn\'{e}strasse 5, 04103 Leipzig, Germany}
\address[add2]{Solid State Physics Division, Bhabha Atomic Research Centre, Mumbai 400 085, India}

\begin{abstract}
We report on the magnetic properties of zinc ferrite thin film deposited on SrTiO$_3$ single crystal using pulsed laser deposition. X-ray diffraction result indicates the highly oriented single phase growth of the film along with the presence of the strain. In comparison to the bulk antiferromagnetic order, the as-deposited film has been found to exhibit ferrimagnetic ordering with a coercive field of 1140~Oe at 5~K. A broad maximum, at $\approx$105~K, observed in zero-field cooled magnetization curve indicates the wide grain size distribution for the as-deposited film. Reduction in magnetization and blocking temperature has been observed after annealing in both argon as well as oxygen atmospheres, where the variation was found to be dependent on the annealing temperature.
\end{abstract}


\end{frontmatter}


\section{Introduction}
Spinel ferrites, AB$_{2}$O$_{4}$, generally exhibit cubic spinel structure, where oxygen anions reside at fcc lattice sites and cations occupy the tetrahedrally and octahedrally coordinated interstitial sites forming A and B sublattices \citep{Brabers1995}. These materials can have normal, inverse and mixed spinel structures and possess different kind of magnetic characters (ferrimagnetic, antiferromagnetic and paramagnetic) depending on the nature of cations and their distribution among different sites\citep{Brabers1995, SchiesslPotzelKarzelEtAl1996}. The zinc ferrite (ZnFe$_{2}$O$_{4}$) has been proposed to be a candidate for spintronic applications, and various studies have been carried out on its magnetic and electrical properties\citep{SchiesslPotzelKarzelEtAl1996, LorenzBrandtMexnerEtAl2011, StewartFigueroaLopezEtAl2007, OliverHamdehHo1999, ChenSpoddigZiese2008, SultanSingh2009, NakashimaFujitaTanakaEtAl2007, ZviaginKumarLoriteEtAl2016}. Bulk zinc ferrite, in perfect oxygen stoichiometry, is known to exhibit normal spinel structure with all Zn$^{2+}$ and Fe$^{3+}$ ions occupying the tetrahedral and octahedral sites, respectively and exhibits antiferromagnetic ordering below 10.5 K \citep{SchiesslPotzelKarzelEtAl1996}. However, nanoparticles \citep{StewartFigueroaLopezEtAl2007, OliverHamdehHo1999} and thin films of ZnFe$_{2}$O$_{4}$ \citep{ChenSpoddigZiese2008, SultanSingh2009, NakashimaFujitaTanakaEtAl2007, ZviaginKumarLoriteEtAl2016} are reported to have the ferrimagnetic order. This is normally attributed to the placement of iron and zinc ions at both the sites, altering the spinel structure from normal to mixed state and inducing strong negative $J_{AB}$ interactions between iron ions \citep{Lotgering1966, KamiyamaHanedaSatoEtAl1992, DormannNogues1990}. Apart from this redistribution of cations, the oxygen vacancy concentration is also believed to play a crucial role in controlling the magnetic properties of zinc ferrite. Greater magnetic response has been observed for the samples prepared under low oxygen partial pressures \citep{ChenSpoddigZiese2008, TorresPasquevichZelisEtAl2014}. In fact, it is still not established whether the observed ferrimagnetism is only related to cation (partial) inversion and/or oxygen vacancies. In this paper, we present magnetic studies of ZnFe$_{2}$O$_{4}$ thin film deposited on SrTiO$_{3}$ (100) substrate, as zinc ferrite is known to grow epitaxially on these substrates \cite{LorenzBrandtMexnerEtAl2011, ZviaginKumarLoriteEtAl2016}. Also, SrTiO$_{3}$ single crystals do not have magnetic impurities and annealing at the studied temperatures is not known to affect their magnetic properties. Hence, SrTiO$_{3}$ is a good choice as a substrate to study the changes in the magnetic properties of the deposited films.

\section{Experimental}\label{experimental details}

Pulsed laser deposition (KrF excimer laser: $\lambda$ = 248~nm) was used to grow a thin film ($\sim$20~nm thick) of zinc ferrite on SrTiO$_{3}$ (100) substrate kept at 773~K and under a relatively low oxygen partial pressure of $6 \times 10^{-5}$ Torr. Prior to deposition, the
base pressure of the chamber was reduced to $\sim7.5 \times 10^{-8}$~Torr. The crystalline structure of the deposited film was studied by x-ray diffraction (XRD) $\theta-2\theta$ scan, performed using the Phillips X'Pert Bragg-Brentano diffractometer with Cu $K_{\alpha}$ radiations. Temperature and field dependent magnetization measurements were carried out using a MPMS-7 superconducting quantum interference device magnetometer from Quantum Design. Zero-field cooled (ZFC) and field cooled (FC) curves were recorded with an applied field of 1000 Oe. To study the effect of annealing environment and temperature on magnetic properties, the as-deposited film was first annealed in argon at 773, 823, and 873~K and subsequently in oxygen at similar temperatures. The magnetization measurements were carried out after each step.

\section{Results and Discussions}\label{Results}

\begin{figure}[h]
\includegraphics[width=0.9\linewidth]{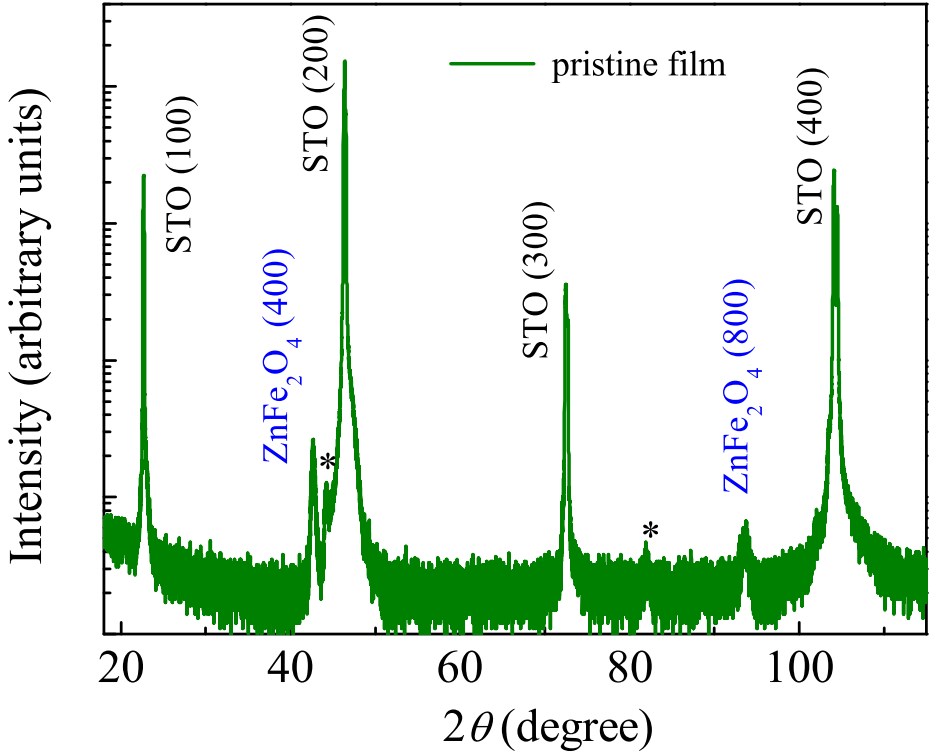}
\caption{\label{XRD} X-ray diffraction scans of pristine zinc ferrite film grown under an oxygen partial pressure of $6 \times 10^{-5}$ Torr. Peaks marked with "*" correspond to the tungsten and related compounds from xrd tube.}
\vspace{-5 mm}
\end{figure}

X-ray diffraction $\theta-2\theta$ pattern of the pristine zinc ferrite film is shown in Fig.~\ref{XRD}, which indicates the epitaxial growth of the film on SrTiO$_{3}$ without presence of any kind of secondary phase. The out of plane lattice parameter has been calculated to be $\sim$8.47 \AA. It should be noted that the lattice parameter is higher than that of bulk value (8.44 \AA)\citep{SpencerSchroeer1974}, which may be due to the substrate induced strain during the thin film growth. Average crystallite size in the deposited film is $\sim$20~nm, as calculated using the Scherrer's formula. No change has been observed in the lattice parameter and crystallinity of the film, within the resolution limit of the instrument, even after the final annealing in the oxygen (not shown here).

\begin{figure}[h]
\includegraphics[width=0.9\linewidth]{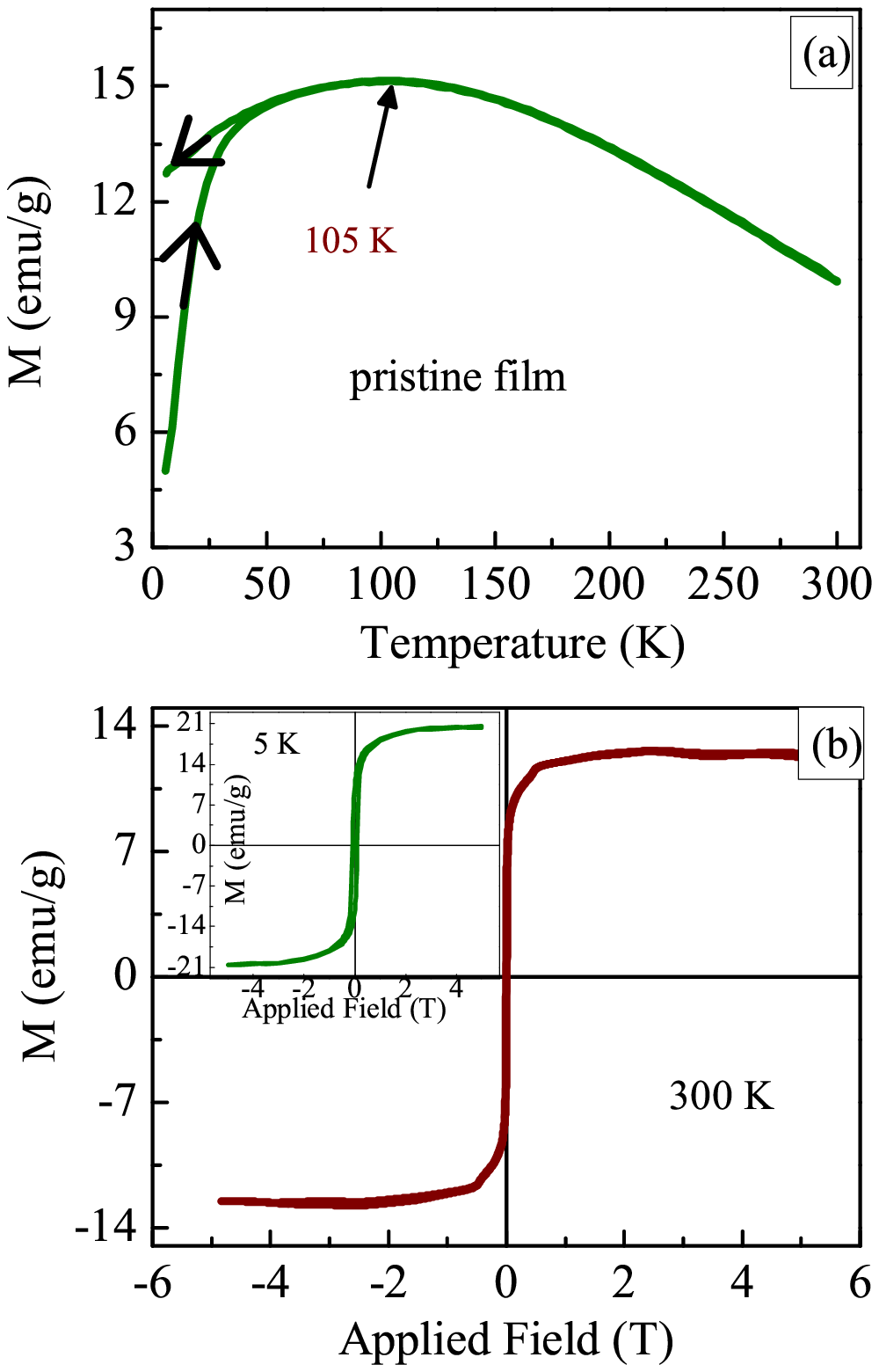}
\caption{\label{MAG1} Variation in mass magnetization (a) with temperature (ZFC-FC) while applying a field of 1000 Oe and (b) with field at 300 K for the as-deposited zinc ferrite film. Inset: mass magnetization vs field curve at 5 K.}
\vspace{-5 mm}
\end{figure}

Field dependent magnetization at 5 K and 300 K, and ZFC-FC curves obtained for the pristine sample are shown in Fig.~\ref{MAG1}. The data has been presented here after subtracting the diamagnetic contribution from the substrate. In comparison to the antiferromangetic order found in bulk zinc ferrite grown under optimal conditions, the observation of S-shaped M-H curves in the as-deposited film indicate the existence of ferrimagnetic order. Such kind of magnetic behaviour has also been observed previously for the thin films of zinc ferrite \citep{ChenSpoddigZiese2008, SultanSingh2009, NakashimaFujitaTanakaEtAl2007,  ZviaginKumarLoriteEtAl2016}. Saturation magnetization ($M_{s}$) at 5~K has been found to be around 20.5~emu/g, equivalent to a magnetic moment of $\sim$0.44 $\mu_{b}$/Fe, with a coercive field of $\approx$ 1140~Oe. At 300~K, along with negligible coercive field, $M_{s}$ is around 12.1 emu/g corresponding to a magnetic moment of $\sim$0.26 $\mu_{b}$/Fe (see Fig.~\ref{MAG1}(b)). In comparison to 300 K, film is magnetically much harder at low temperature, indicating the presence of some sort of blocking mechanism, which is also supported by the irreversibility in the ZFC-FC curves at lower temperatures. Apart from this, a broad maximum is present in the ZFC curve. Here, the magnetization first increases with the temperature and after $\sim$105~K, known as blocking temperature ($T_{B}$), it reduces suggesting a wide particle size distribution for the as-deposited film.
\begin{figure*}
\includegraphics[width=\textwidth]{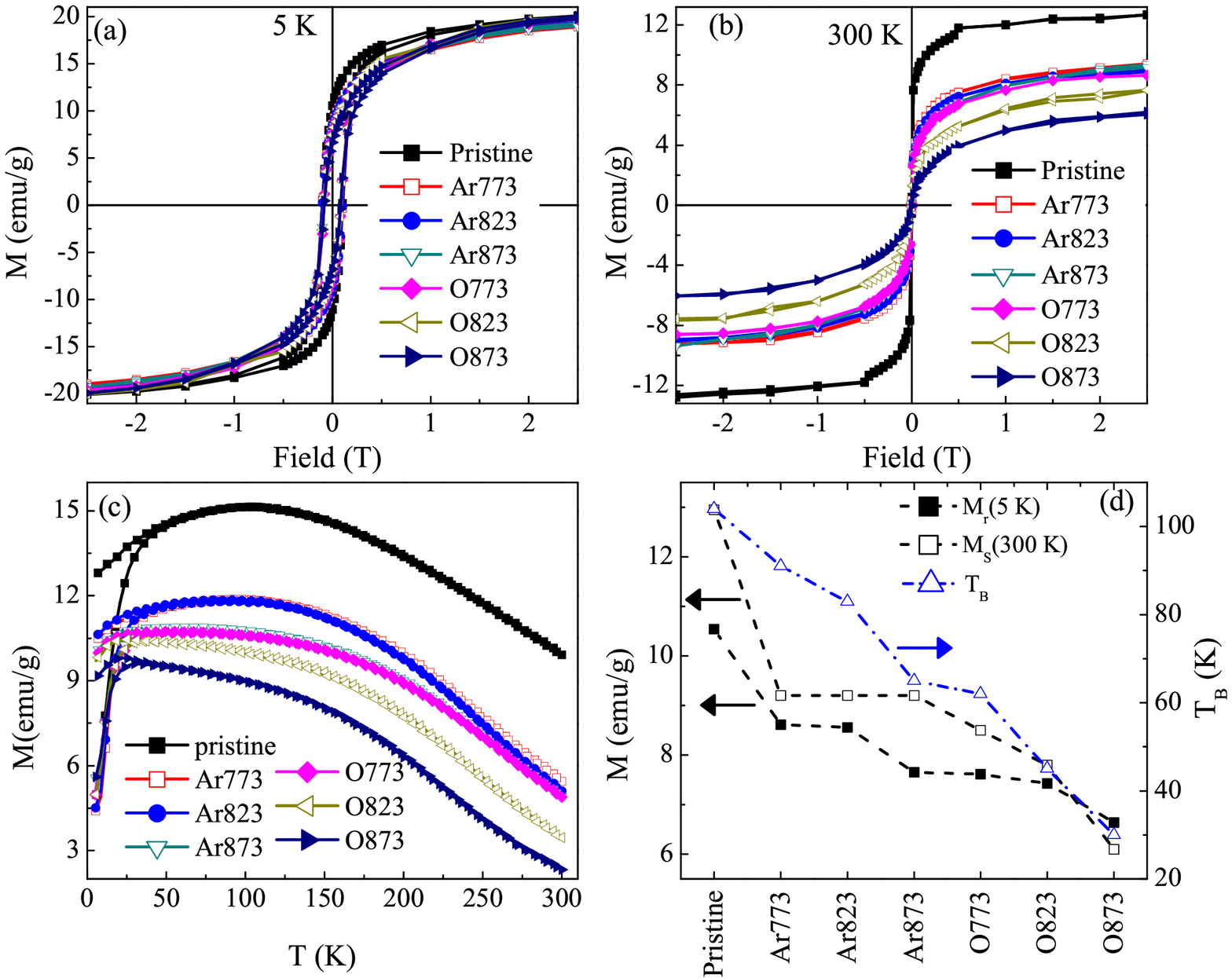}
\caption{\label{cMAG1} Variation in mass magnetization with field at (a) 5 K and (b) 300 K ; (c) ZFC-FC curves recorded with an applied field of 1000 Oe and, (d) extracted residual magnetization ($M_{r}$) at 5 K, spontaneous magnetization ($M_{s}$) at 300 K and blocking temperature($T_{B}$) for pristine zinc ferrite film and after annealing under various conditions.}
\vspace{-6 mm}
\end{figure*}

It has previously been observed that the oxygen vacancies play a major role in the magnetic properties of zinc ferrite, with larger magnetic reponse for films grown under lower oxygen partial pressures \citep{ChenSpoddigZiese2008}. Hence, at first glance, it seems that the observed magnetization can be explained on the basis of oxygen vacancies, as the film is grown at quite low oxygen partial pressure of $6 \times 10^{-5}$ Torr. To verify this assumption and to avoid any removal of oxygen vacancies, we first annealed the film  at 773, 823 and 873~K in the argon gas. After these treatments, we performed annealing in oxygen environment to remove the oxygen vacancies. Magnetization measurements were carried out after each treatment. It is important to mention here that in each case the annealing temperature/atmosphere was changed only after getting the saturation effect at the current temperature, which was studied by the magnetization measurements with multiple treatments under similar conditions. In most cases, saturation effect was achieved after 4 hours of annealing. Interestingly, with annealing in the argon atmosphere, magnetization of the sample has been found to be reduced with the rise in the temperature. The saturation magnetization ($M_{s}$) is only affected at 300 K, while at 5 K remanence ($M_{r}$) reduces after annealing (see Fig.~\ref{cMAG1}(a)). ZFC-FC curves still exhibit the irreversibility at low temperatures; however, the blocking temperature decreases after each treatment (shown separately in Fig.~\ref{cMAG1}(d)). Such effect can not be explained on the basis of oxygen vacancies as the argon-annealing is not supposed to remove these vacancies. However, some kind of structural changes can also take place during annealing at higher temperatures. In fact, in one of our recent article, we have shown that the substrate temperature plays a major role on occupancy of tetrahedral sites by iron ions thereby affecting the magnetic properties \cite{ZviaginKumarLoriteEtAl2016}. It was found that with the rise in the substrate temperature, zinc ferrite tends to have normal spinel structure with reduced magnetization. Hence, under argon-annealing, reduction in magnetization for studied sample with rise in temperature can also be explained on the basis of stabilization of normal spinel structure.

After argon-annealing, the film was annealed in oxygen and obtained results are plotted in Fig.~\ref{cMAG1}. Magnetization was reduced with annealing in oxygen as well, again showing the major changes in the saturation values at 300 K. Also, the blocking temperature decreases further with dependence on temperature (see Fig. 3(d)), along with the appearance of irreversibility in ZFC-FC curves even after the final annealing step. As the annealing under oxygen was performed at temperatures not higher than that used for argon-annealing, hence, we can ignore the possibility of further redistribution of iron ions at different sites. Thus, these changes can be easily explained on the basis of removal of oxygen vacancies, which is consistent with the reports on zinc ferrite films where less magnetization has been observed for the samples prepared under higher oxygen partial pressures \citep{ChenSpoddigZiese2008}. Hence, on the basis of our observations, we can say that the magnetic properties of zinc ferrite films  are affected both by structural changes through cation inversion as well as oxygen vacancies, and can be controlled by the annealing temperature and the atmosphere. It is important to mention here that similar kind of magnetization study was also performed on other comparable zinc ferrite films, which exhibited similar results. However, for the sake of brevity this data is not presented here.

\section{Conclusions}\label{cocl}
In summary, we have studied the magnetic properties of highly oriented zinc ferrite thin film grown on SrTiO$_{3}$ single crystals and their dependence on annealing in the temperature range of 773-873~K and under different atmospheres. Pristine film exhibits ferrimagnetic order and a broad zero field cooled magnetization curve. Magnetization has been found to be reduced after annealing in argon and oxygen with larger changes observed for measurements at 300 K. Our observations demonstrate that the magnetic properties of zinc ferrite films can be controlled by varying the annealing temperature under different environments.

\section*{Acknowledgments}\label{acknow}
This work was supported  by the DFG within the Collaborative
Research Center (SFB 762) ``Functionality of Oxide Interfaces". One of the authors (YK) would like to thank Department of Science and Technology (DST), India for the financial support. We also acknowledge Annette Setzer for the technical support.


\begin{thebibliography}{10}
\expandafter\ifx\csname url\endcsname\relax
  \def\url#1{\texttt{#1}}\fi
\expandafter\ifx\csname urlprefix\endcsname\relax\def\urlprefix{URL }\fi
\expandafter\ifx\csname href\endcsname\relax
  \def\href#1#2{#2} \def\path#1{#1}\fi

\bibitem{Brabers1995}
V.~Brabers, \href{http://dx.doi.org/10.1016/S1567-2719(05)80032-0}{Chapter 3
  progress in spinel ferrite research}, in: Handbook of Magnetic Materials,
  Elsevier {BV}, 1995, pp. 189--324.



\bibitem{SchiesslPotzelKarzelEtAl1996}
W.~Schiessl, W.~Potzel, H.~Karzel, M.~Steiner, G.~M. Kalvius, A.~Martin, M.~K.
  Krause, I.~Halevy, J.~Gal, W.~Schäfer, G.~Will, M.~Hillberg, R.~Wäppling,
  \href{http://dx.doi.org/10.1103/PhysRevB.53.9143}{Magnetic properties of the
  {ZnFe$_{2}$O$_{4}$} spinel}, Phys. Rev. B 53~(14) (1996) 9143--9152.


\bibitem{LorenzBrandtMexnerEtAl2011}
M.~Lorenz, M.~Brandt, K.~Mexner, K.~Brachwitz, M.~Ziese, P.~Esquinazi,
  H.~Hochmuth, M.~Grundmann,
  \href{http://dx.doi.org/10.1002/pssr.201105359}{Ferrimagnetic
  {ZnFe$_{2}$O$_{4}$} thin films on {SrTiO$_{3}$} single crystals with highly
  tunable electrical conductivity}, Phys. Status Solidi {RRL} 5~(12) (2011)
  438--440.

\bibitem{StewartFigueroaLopezEtAl2007}
S.~J. Stewart, S.~J.~A. Figueroa, J.~M.~R. L{\'{o}}pez, S.~G. Marchetti, J.~F.
  Bengoa, R.~J. Prado, F.~G. Requejo,
  \href{http://dx.doi.org/10.1103/PhysRevB.75.073408}{Cationic exchange in
  nanosized {ZnFe$_{2}$O$_{4}$} spinel revealed by experimental and simulated
  near-edge absorption structure}, Phys. Rev. B 75~(7).


\bibitem{OliverHamdehHo1999}
S.~A. Oliver, H.~H. Hamdeh, J.~C. Ho,
  \href{http://dx.doi.org/10.1103/PhysRevB.60.3400}{Localized spin canting in
  partially inverted {ZnFe$_{2}$O$_{4}$} fine powders}, Phys. Rev. B 60~(5)
  (1999) 3400--3405.

\bibitem{ChenSpoddigZiese2008}
Y.~F. Chen, D.~Spoddig, M.~Ziese,
  \href{http://dx.doi.org/10.1088/0022-3727/41/20/205004}{Epitaxial thin film
  {ZnFe$_{2}$O$_{4}$} : a semi-transparent magnetic semiconductor with high
  curie temperature}, Journal of Physics D: Applied Physics 41~(20) (2008)
  205004.

\bibitem{SultanSingh2009}
M.~Sultan, R.~Singh, \href{http://dx.doi.org/10.1063/1.3072381}{Magnetic and
  optical properties of rf-sputtered zinc ferrite thin films}, J. Appl. Phys.
  105~(7) (2009) 07A512.

\bibitem{NakashimaFujitaTanakaEtAl2007}
S.~Nakashima, K.~Fujita, K.~Tanaka, K.~Hirao, T.~Yamamoto, I.~Tanaka,
  \href{http://dx.doi.org/10.1103/PhysRevB.75.174443}{First-principles {XANES}
  simulations of spinel zinc ferrite with a disordered cation distribution},
  Phys. Rev. B 75~(17).


\bibitem{ZviaginKumarLoriteEtAl2016}
V.~Zviagin, Y.~Kumar, I.~Lorite, P.~Esquinazi, M.~Grundmann, R.~Schmidt-Grund,
  \href{http://dx.doi.org/10.1063/1.4944898}{Ellipsometric investigation of
  {ZnFe$_{2}$O$_{4}$} thin films in relation to magnetic properties}, Appl.
  Phys. Lett. 108~(13) (2016) 131901.

\bibitem{Lotgering1966}
F.~Lotgering, \href{http://dx.doi.org/10.1016/0022-3697(66)90175-2}{The
  influence of fe$^{3+}$ ions at tetrahedral sites on the magnetic properties
  of {ZnFe$_{2}$O$_{4}$}}, Journal of Physics and Chemistry of Solids 27~(1)
  (1966) 139--145.

\bibitem{KamiyamaHanedaSatoEtAl1992}
T.~Kamiyama, K.~Haneda, T.~Sato, S.~Ikeda, H.~Asano,
  \href{http://dx.doi.org/10.1016/0038-1098(92)90412-3}{Cation distribution in
  {ZnFe$_{2}$O$_{4}$} fine particles studied by neutron powder diffraction},
  Solid State Communications 81~(7) (1992) 563--566.

\bibitem{DormannNogues1990}
J.~L. Dormann, M.~Nogues,
  \href{http://dx.doi.org/10.1088/0953-8984/2/5/014}{Magnetic structures in
  substituted ferrites}, Journal of Physics: Condensed Matter 2~(5) (1990)
  1223--1237.


\bibitem{TorresPasquevichZelisEtAl2014}
C.~E.~R. Torres, G.~A. Pasquevich, P.~M. Z{\'{e}}lis, F.~Golmar, S.~P. Heluani,
  S.~K. Nayak, W.~A. Adeagbo, W.~Hergert, M.~Hoffmann, A.~Ernst, P.~Esquinazi,
  S.~J. Stewart,
  \href{http://dx.doi.org/10.1103/PhysRevB.89.104411}{Oxygen-vacancy-induced
  local ferromagnetism as a driving mechanism in enhancing the magnetic
  response of ferrites}, Phys. Rev. B 89~(10).

\bibitem{SpencerSchroeer1974}
C.~D. Spencer, D.~Schroeer,
  \href{http://dx.doi.org/10.1103/PhysRevB.9.3658}{Mössbauer study of several
  cobalt spinels using {Co$^{57}$ and Fe$^{57}$}}, Phys. Rev. B 9~(9) (1974)
  3658--3665.

\end{thebibliography}
\end{document}